\documentclass[english,twocolumn,prl,preprintnumbers,amsmath,amssymb,superscriptaddress]{revtex4}
\usepackage{verbatim}
\usepackage{graphicx}
\usepackage{amssymb}
\usepackage{babel}
\makeatother

\newcommand{\bra}[1]{\left\langle {#1} \right|}
\newcommand{\ket}[1]{\left|  #1 \right\rangle}

\newcommand{\unitv}[1]{\mathbf{\hat{#1}}}
\newcommand{\topt}[0]{t_\mathrm{op}}
\newcommand{\tw}[0]{t_\mathrm{w}}
\newcommand{\tr}[0]{t_\mathrm{r}}
\newcommand{\tauL}[0]{\tau_L}

\begin{document}

\preprint{XXX}

\title{Heralded single-magnon quantum memory for photon polarization states}

\author{Haruka Tanji}
\affiliation{Department of Physics, Harvard University, Cambridge,
Massachusetts 02138, USA}
\affiliation{Department of Physics,
MIT-Harvard Center for Ultracold Atoms, and Research Laboratory of
Electronics, Massachusetts Institute of Technology, Cambridge,
Massachusetts 02139, USA}
\author{Saikat Ghosh}
\affiliation{Department of Physics, MIT-Harvard Center for
Ultracold Atoms, and Research Laboratory of Electronics,
Massachusetts Institute of Technology, Cambridge, Massachusetts
02139, USA}
\author{Jonathan Simon}
\affiliation{Department of Physics, Harvard University, Cambridge,
Massachusetts 02138, USA}
\affiliation{Department of Physics,
MIT-Harvard Center for Ultracold Atoms, and Research Laboratory of
Electronics, Massachusetts Institute of Technology, Cambridge,
Massachusetts 02139, USA}
\author{Benjamin Bloom}
\affiliation{Department of Physics, MIT-Harvard Center for
Ultracold Atoms, and Research Laboratory of Electronics,
Massachusetts Institute of Technology, Cambridge, Massachusetts
02139, USA}
\author{Vladan Vuleti\'{c}}
\affiliation{Department of Physics, MIT-Harvard Center for Ultracold Atoms, and
Research Laboratory of Electronics, Massachusetts Institute of Technology,
Cambridge, Massachusetts 02139, USA}

\date{\today}

\begin{abstract}
We demonstrate a heralded quantum memory based on mapping of a
photon polarization state onto a single collective-spin excitation
(magnon) shared between two spatially overlapped atomic ensembles. The polarization fidelity is measured by quantum state tomography to be above $90(2)\%$ for any input polarization, which far exceeds the classical limit of $\frac{2}{3}$. 
The process also constitutes a quantum non-destructive probe that detects and regenerates a photon without touching its --- potentially undetermined --- polarization.
\end{abstract}


\maketitle

The power of quantum communication can be boosted by quantum memories \cite{Duan01,Kielpinski01,Chou04,Matsukevich04,Schoelkopf2004,Chaneliere05,Eisaman05,Black05a,Petta05,Thompson06,Choi08}
that can receive, store, and release a quantum state typically carried
by a photon. The advantages memories offer, however, are often thwarted
by photon losses \cite{Duan01,Simon07c,Jiang07,Chen07}. Such
unpredictable failure may be largely remedied by a heralding
feature that announces photon arrival and successful storage
without destroying or revealing the stored quantum state.
Heralded storage may thus advance long-distance quantum communication \cite{Duan01},
linear-optics quantum computing \cite{Knill01}, or schemes aimed at
breaking quantum encryption \cite{Sanders00}.

Quantum state storage has been investigated in various
systems \cite{Kielpinski01,Petta05,Schoelkopf2004,Chou04,Matsukevich04,Chaneliere05,Eisaman05,Black05a,Thompson06}.
Atomic-ensemble quantum memories have been pursued both for continuous
variables of electromagnetic fields \cite{Julsgaard01,Julsgaard04}, and for quantized photonic excitations \cite{Chou04,Matsukevich04,Chaneliere05,Eisaman05,Black05a,Thompson06,Choi08}.
In an elegant experiment, Julsgaard \textit{et.\ al.}\
\cite{Julsgaard04} mapped the quadrature variables of a weak coherent
field onto an atomic ensemble through a field
measurement and subsequent feedback onto the ensemble.
Other advancements towards a continuous-variable memory include
the recent demonstration of storage and retrieval of squeezed
vacuum \cite{Honda08,Appel08}.

Much progress has been made in the storage and retrieval of
individual photons. Early work demonstrated capture and release of
single photons of fixed polarization using electromagnetically induced transparency
\cite{Chaneliere05,Eisaman05}, as well as their adiabatic transfer
between two ensembles via an optical resonator \cite{Simon07b}. Matsukevich and Kuzmich showed that two atomic ensembles can serve
as a two-level system whose state can be prepared by a projective
measurement \cite{Matsukevich04}. Recently, Choi \textit{et.\ al.}\ mapped
photonic entanglement created by a polarizing beamsplitter onto
two ensembles, and later retrieved the photon \cite{Choi08}, realizing unheralded, but
relatively high-efficiency, polarization storage. In
work by Chen \textit{et.\ al.}, a successful Bell measurement between
two photons resulted in probabilistic teleportation of a photon
polarization state onto two atomic ensembles \cite{Chen08}.
This can be viewed as a partially heralded quantum memory, where a two-photon coincidence
between two beams with Poissonian statistics sometimes, but not
always, heralds a successful Bell measurement and teleportation \cite{Chen08}.

In this Letter, we demonstrate a system where a single photon
announces polarization storage in the form of a single
collective-spin excitation (magnon) that is shared between two spatially overlapped
atomic ensembles. The heralded storage
occurs rarely ($p \sim 10^{-6}$ in our non-optimized setup), but when it does, the incident
photon is stored and can later be recreated with good efficiency
($\varepsilon \sim 50\%$) and sub-Poissonian statistics ($g_2
\approx 0.24$), while its polarization state is restored with very
high fidelity ($\mathcal{F}> 90\%$).

\begin{figure}
\includegraphics[width=3.2in,keepaspectratio]{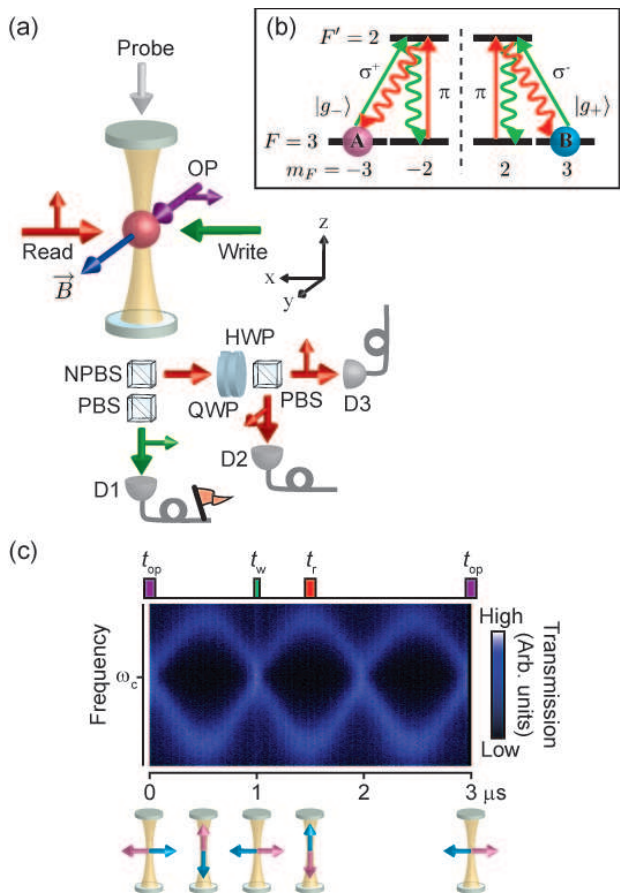}
\caption{(Color online.) (a) Setup. Small arrows indicate beam polarization, OP is the optical pumping beam. NPBS, PBS, QWP, and HWP denote a non-polarizing beamsplitter, a polarizing beamsplitter, a quarter-waveplate, and a half waveplate, respectively. D1, D2, D3 are single-photon counting modules. A static magnetic field produces magnon precession.
(b) Energy levels. Ensembles $A$ and
$B$ are initially prepared in $\ket{g_\mp}\equiv\ket{F=3,m_F=\mp3}$.
The write (green) and the read (red) processes are $\sigma^\pm$-$\pi$ and $\pi$-$\sigma^\pm$
spontaneous Raman transitions, respectively.
(c) Precession of the two macroscopic spins, as measured by cavity transmission spectroscopy, and timing of the optical-pumping ($\topt$), write ($\tw$), and read ($\tr$) processes.}
\label{setup}
\end{figure}

Heralded storage is achieved by means of a spontaneous Raman
process that simultaneously creates a photon of fixed polarization
(that serves as the herald), and a collective spin excitation (magnon) \cite{Duan01,Simon07} that is a copy of the input-beam polarization. To store an arbitrary polarization state
\begin{equation}
\ket{\psi} = \cos\theta\ket{R} + e^{i\phi}\sin\theta\ket{L},
\label{photonstate}
\end{equation}
written as a superposition of two right/left circularly polarized states $\ket{R}$, $\ket{L}$ with two arbitrary angles $\theta,
\phi$, we use two spatially overlapped atomic ensembles $A,B$
inside an optical resonator. The atomic levels are chosen such
that ensemble $A$ ($B$) absorbs only $\ket{R}$ ($\ket{L}$) polarized light, while both can emit a photon of the same polarization ($\pi$) into the resonator on the Raman transition
of interest (Fig. \ref{setup}). The detection of the emitted $\pi$ photon heralds the
mapping of the input polarization state onto a magnon,
but does not provide ``which-path'' information to
distinguish between $A$ and $B$. The heralding also ensures that, even if
the input is a coherent beam, only one magnon is generated between
the two ensembles in the limit of small Raman scattering
probability. The ``write'' process thus projects a polarization state $\ket{\psi}$ onto a magnon superposition state
\begin{equation}
\ket{\psi} \rightarrow \ket{\Psi} = \cos \theta \ket{1}_A\ket{0}_B + e^{i \phi} \sin \theta  \ket{0}_A \ket{1}_B,
\label{magnonstate}
\end{equation}
where $\ket{n}_k$ denotes $n$ magnons in ensemble $k$ ($k=A,B$).
For general input polarization ($\theta \neq 0, \frac{\pi}{2},
\pi$), this process creates an entangled state of the
two ensembles. At a later time, the stored state can be retrieved
on demand as a single photon by utilizing the strong coupling of
the magnon to the resonator mode \cite{Duan01,Simon07} (``read'' process).

The heralding serves to enhance the fidelity of the write process
by announcing successful events. In our present non-optimized setup, the heralding probability $p= \alpha_\perp \eta  q   \approx 10^{-6}$, 
being the product of optical depth perpendicular to the resonator ($\alpha _\perp = 0.01$), 
emission probability into the resonator (single-atom cooperativity $\eta=10^{-3}$), and photon detection efficiency ($q=0.1$), is low. 
Whenever there is a heralding event,
however, a single magnon corresponding to the input-field polarization is stored with high fidelity. The single-photon
nature of the retrieved field is confirmed by a conditional
autocorrelation measurement indicating four-fold suppression of
two-photon events compared to a Poissonian source ($g_2=0.24(5)$).
The heralding process may thus be alternatively viewed as a quantum non-demolition measurement of a single photon \cite{Guerlin07} 
which preserves the polarization, and stores the photon.

The heralded storage is performed with precessing spins \cite{Matsukevich06b} in order to make use of atomic symmetries for 
good polarization fidelity, and resonator emission in both heralding and read processes for mode selection and coupling efficiency. 
We choose a $\pi$ transition for the heralding photon, while any input state is
expressed as a superposition of $\sigma^{\pm}$ polarization (Fig.\ \ref{setup}b).
Given the corresponding atomic angular emission patterns, we then need to rotate the atomic spin direction by $90^\circ$ between the heralding and the readout.
This is achieved with a magnetic field of $1.4\ \mathrm G$ that induces Larmor spin
precession with a period of $\tau_L=2\ \mathrm{\mu s}$
(Fig.\ \ref{setup}a, \ref{setup}d), enabling us to access the same magnon with different light polarizations at different times. Note that a spatially
homogeneous magnetic field maintains the inter-atomic coherence,
and does not affect the magnon momentum, or equivalently, the
phase matching condition for the read process \cite{Black05a}.

An ensemble of $N_0 \sim3\times10^{4}$ cesium atoms at a temperature of $30\ \mu\mathrm K$ is loaded 
from a magneto-optical trap into a far-detuned (trap wavelength $\lambda_t = 1064\ \mathrm{nm}$) 
one-dimensional optical lattice overlapped with the mode of a medium-finesse ($f=140$) optical resonator at the waist. 
We prepare a subset $N$ of the atoms in the $6S_{1/2}$, $F=3$ hyperfine ground state with a resonant optical depth $N\eta\approx 16$. 
Ensembles $A$ and $B$ each consist of approximately $N/2$ atoms, optically pumped into hyperfine and magnetic sublevels 
$\ket{g_{\pm}} \equiv \ket{6S_{1/2},F=3,m_F=\pm 3}$, respectively, in the rotating frame. 
(The quantization axis is defined to rotate with the atomic spins and coincide with the propagation direction of the write beam at 
the optical pumping time $\topt=0$.) Optical pumping in this frame is achieved by periodic application of a short ($100\ \mathrm{ns} \ll \tau_L$), 
linearly ($\mathbf{\hat x}$-) polarized optical pumping pulse, resonant with the $6S_{1/2}, F=3\rightarrow 6P_{3/2}, F^\prime=2$ transition. The ensembles
$A,B$ thus form macroscopic spins that point in opposite directions, and Larmor precess with a period of $\tau_L$ in the $x$-$z$ plane (Fig.\ \ref{setup}c). 
We choose a pumping period of $1.5 \tau_L$, such that the ensembles are interchanged at every trial which removes any systematic effects 
due to a population imbalance between $\ket{g_\pm}$.

The atomic-spin precession and the efficiency of the optical pumping may be monitored by sending a weak, 
linearly ($\mathbf{\hat x}$-) polarized probe beam through the resonator.
In a coordinate system rotating with the atomic spin, the probe
beam polarization changes periodically with time. When the
probe beam as seen by the atoms is $\pi$-polarized, the states
$\ket{g_\pm}$ do not couple to the probe light on the chosen
transition $F=3 \rightarrow F^\prime=2$ (see Fig.\ \ref{setup}c), and the otherwise observable atom-induced
splitting (Rabi splitting) of the cavity resonance \cite{Zhu90}
disappears. The sinusoidal variation of the Rabi splitting (Fig.\
\ref{setup}c) is observed only for a polarized sample. By
maximizing the contrast of the oscillation, we optically pump more than 99\% of the atoms in the $F=3$ hyperfine manifold into either of the $\ket{g_{\pm}}$ sublevels.

The photon storage and readout processes are timed to match the
sample precession (Fig.\ \ref{setup}c). 
A sequence of optical-pump, write and read pulses is applied once every $1.5\tau_L = 3\ \mu s$ for $30\ \mathrm{ms}$, corresponding to a total of $10^4$ trials. 
The set of sequences is repeated at $\sim 0.5\ \mathrm{Hz}$ to allow for recooling of the sample in between.
The write beam whose polarization, as set by a variable retarder and a half-wave plate, is to be stored, 
propagates along the $\mathbf{\hat{x}}$ direction, and is tuned to $F=3\rightarrow F^\prime=2$ atomic transition. 
It is pulsed on for $50\ \mathrm{ns}\ll\tau_L$ at $\tw=\tauL/2 =1 \mu\mathrm s$, when the
macroscopic spins are aligned along $\mp \unitv{x}$. At this time,
$\ket{R}$ and $\ket{L}$ correspond to $\sigma^\pm$ transitions
along the quantization axis $\unitv{x}$, such that $A$ and $B$ can
absorb only $\ket{R}$ and $\ket{L}$ photons, respectively (Fig.\ \ref{setup}b).
The write intensity is adjusted such that
much less than one photon is scattered into the resonator mode per pulse.
For equal populations in $A$ and $B$, a $\pi$-polarized photon
originating from a spontaneous $\sigma^{\pm}$-$\pi$ (absorbing a $\sigma^\pm$ photon and emitting a $\pi$ photon) Raman process
has the same probability for having been emitted by either
ensemble. Thus, it does not provide any ``which-path''
information, and, if detected, serves as a heralding photon that
announces the storage of a (not revealed) polarization state $\ket{\psi}$ as a magnon
$\ket{\Psi}$.
The heralding photon is detected by detector D1 (Fig.\ \ref{setup}a).

At $\tr = \tw+\tauL/4 = 1.5\ \mu\mathrm s$, when
the atomic spins point along the resonator axis $\pm
\unitv{z}$, the read beam, linearly polarized along $\unitv{z}$ and tuned to $F=3\rightarrow F^\prime=2$ transition, is applied for $100\ \mathrm{ns}\ll\tauL$. The read beam excites the atoms on
a $\pi$ transition, such that collectively enhanced \cite{Duan01}
$\pi$-$\sigma^{\pm}$ Raman scattering maps the magnon state onto a
single-photon polarization state. If the relative populations, $|\cos \theta|^2$, $|\sin \theta|^2$, and the relative phase $\phi$ of the magnons in ensembles $A,B$ are
preserved between the write and read processes (Eq.\
\ref{magnonstate}), the polarization of the regenerated photon should be
a faithful copy of the write beam polarization.

\begin{figure}
\includegraphics[width=3.4in,keepaspectratio]{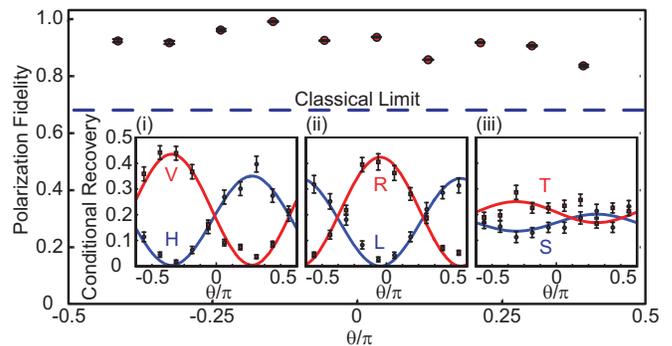}
\caption{(Color online.) Polarization fidelity of the stored photon as
a function of $\theta$ for $\phi=0$ (Eq.\ \ref{photonstate}).
The dashed line indicates the classical limit of $2/3$.
Insets (i)-(iii): The results of projection measurements
of the output field in three mutually-orthogonal bases, H-V, L-R,
and S-T. The solid curves are a simultaneous fit for all sixty data points. The
error bars represent statistical errors due to finite detection
counts. No backgrounds have been subtracted.} \label{projection}
\end{figure}

To investigate the quality of the heralded polarization memory, we evaluate
the polarization fidelity of the retrieved single photon with respect
to the input state. We determine the density matrix $\rho_\mathrm{meas}$ of the output
polarization by measuring the projection onto three polarization bases \cite{James01}: $\frac{1}{\sqrt
2}\left(\ket{L}\pm\ket{R}\right)$ (H-V), $\ket{L}$ and $\ket{R}$
(L-R), and $\frac{1}{\sqrt 2}\left(\ket{L} \pm i \ket{R} \right)$ (S-T).
As the phase $\theta$ of the input state Eq.\
(\ref{photonstate}) is varied, the projection onto those bases
displays a sinusoidal variation as expected (inset of Fig.\
\ref{projection}), confirming that the system behaves linearly.

\begin{figure}
\includegraphics[width=3.4in,keepaspectratio]{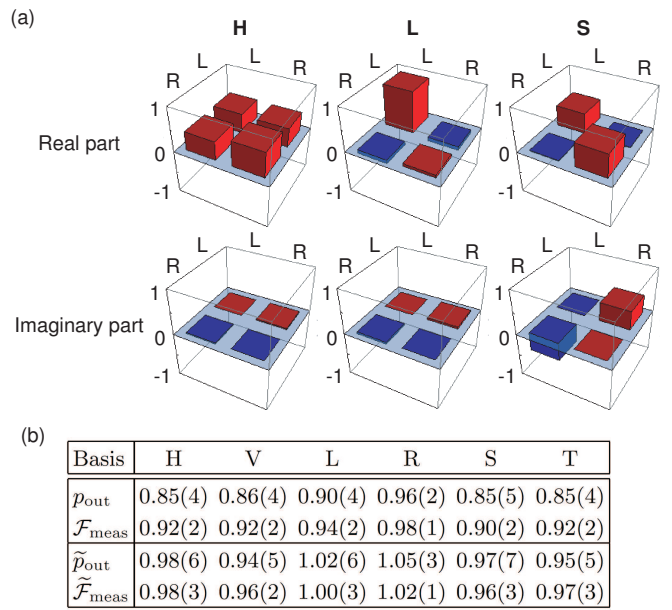}\\
\caption{(Color online.) (a) Density matrices $\rho_\mathrm{meas}$ of the retrieved single photons for fiducial input states $H$, $L$ and $S$. 
(b) The measured degrees of
polarization ($p_\mathrm{out}$) and fidelities ($\mathcal{F}$) of the retrieved single photons for the six fiducial input states. 
The symbols with tilde denote the values with the independently measured photon background subtracted.
} \label{tomography}
\end{figure}

The polarization fidelities $\mathcal{F}$ of the
retrieved single photons for the ten states shown in Fig.\
\ref{projection} as well as for the six fiducial input states, H,
V, L, R, S, and T are evaluated from the density matrices $\rho_\mathrm{meas}$
(some of which are shown in Fig.\ \ref{tomography}a) as $\mathcal{F} =
\mathrm{Tr}( \rho_\mathrm{meas} \ket{\psi}\bra{\psi})$, where $\ket{\psi}$ is
the input state in Eq.\ (\ref{photonstate}). Fig. \ref{projection} shows that $\mathcal{F}$ is close to unity with no notable dependence on 
the zenith angle $\theta$, and we have verified separately that the same is true for the azimuth angle $\phi$. 
In particular, for  any of the six fiducial states the measured fidelity $\mathcal{F}$ without any background subtraction is 
significantly above the classical limit of $2/3$ for state-independent storage (Fig.\ \ref{tomography}b).
If we subtract the independently measured
photonic backgrounds present during the readout, the degrees of
polarization $\widetilde p_\mathrm{out}$, as well as the
fidelities $\widetilde{F}$, approach unity (Fig.\
\ref{tomography}b), indicating that the polarization fidelity inherent in the magnon storage even exceeds that displayed in Fig.\ \ref{projection}.

The major source of photon backgrounds is the finite Larmor precession of $0.3\ \mathrm{rad}$ during the 100-ns read process. 
The read pump beam acquires a small admixture of $\sigma^{\pm}$ component in the frame precessing with the atomic spin (see Fig.\ \ref{setup}b), 
which results in strong photon scattering by atoms in $\ket{g_{\pm}}$ into the resonator. 
These backgrounds deteriorate not only the stored-polarization fidelity, but also the single-photon character of the retrieved field, i.e., 
increase the autocorrelation function. 
These limitations can be overcome by slowing down the Larmor precession, currently limited by the magnon 
Doppler decoherence time of a few microseconds \cite{Simon07}. Implementation of a tightly confining 
three-dimensional optical lattice is expected to substantially reduce the decoherence, and increase the storage time.


Finally, we estimate the degree of entanglement present between samples $A$ and $B$ during storage. 
The amount of entanglement may be quantified by the concurrence $\mathcal{C}$ \cite{Wootters98}, where $1\ge C>0$ indicates entanglement.
The concurrence of an atomic state $\mathcal{C}$ is bounded by that of the corresponding photonic state $\mathcal{C}_\mathrm{ph}$ 
as discussed in the Supplementary Information of Ref.\ \cite{Simon07b}. $\mathcal{C}_\mathrm{ph}$ is observed to vary with zenith angle $\theta$, as expected, with a maximum value of $\mathcal{C}_\mathrm{ph}=0.034(4)$ for the $\ket{H}=\frac{1}{\sqrt 2}(\ket{L}+\ket{R})$ state.

To conclude, we have demonstrated a heralded memory for photon
polarization states with an average fidelity of $0.93(5)$. The low
success probability (currently $\sim10^{-6}$ per trial, including
detection efficiency) may be improved upon by a dipole trap and a
modified resonator which will realistically increase the
transverse optical depth to $1$ and the single-atom cooperativity
in the cavity mode to $0.1$, respectively.
The success
probability and the effective success rate will then be $\gtrsim 1\%$ and
$200\ \mathrm s^{-1}$, respectively. The retrieved photons in this scheme have controllable
waveforms, and can easily be interfered with one another with high
fringe contrast because of their narrow, nearly Fourier-limited
bandwidth \cite{Thompson06}. This is a crucial feature for any
quantum information application. Furthermore, by applying this
scheme to photons of undetermined polarization from a
source of entangled-photon pairs \cite{Wilk07},
it should be possible to realize a \emph{heralded} source of high-quality
entangled-photon pairs for various tasks in quantum information
processing.

We gratefully acknowledge support by the
NSF and DARPA. J.S. thanks the NDSEG and NSF for support.

\end{document}